\documentstyle[12pt,natbib,psfig]{article}

\begin{document}

{\bf
What can we infer about the underlying physics
from burst distributions observed in an RMHD simulation?\\

N. W. Watkins$^{1}$, S. Oughton$^{2}$, M. P. Freeman$^{1}$ \\

$^{1}$British Antarctic Survey (NERC), High Cross, Madingley Road,
Cambridge, CB3 0ET, UK\\

$^{2}$Department of Mathematics, University College London, Gower St,
London WC1E 6BT, UK\\


\abstract{We determine that the sizes of bursts in mean-square
current density in a reduced magnetohydrodynamic (RMHD)
simulation follow a power-law probability density function (PDF).
The PDFs for burst durations and waiting time between bursts
are clearly not exponential and could also be power-law.
This suffices to distinguish their behaviour from the original
Bak et al.\ sandpile model which had exponential waiting time PDFs.
However, it is not sufficient to distinguish between turbulence, some
other SOC-like models, and other red noise sources.}

\section{Introduction}

The widespread occurrence of both self-affine time series with
``$1/f$'' power spectra and spatial fractals in nature led
      \cite{BakEA87,BakEA88} (BTW)
to propose the hypothesis of self-organised criticality (SOC)
    \citep{BakSOC,JensenSOC,SornetteSOC}
as their common origin. Their proposal was based on the
demonstration of a  ``sandpile'' cellular automaton (see also the
earlier work
of \cite{Katz86}) which appeared to be
attracted from arbitrary initial conditions (``self-organisation'') to
a critical state characterised by fluctuations on all scales in the
energy released by the system (``criticality'').
Power-law probability
density functions (PDFs) for the sizes and durations of energy bursts
were the main observed signatures of criticality, and were tested by
finite-size scaling of the PDFs with system size \citep{CardySR}.

One of the original applications proposed by BTW for their idea was
fully developed turbulence, in view of the scaling behaviour of such
systems, and the intermittency of their energy
dissipation. Furthermore, intermittent turbulence has been an
inspiration for later ``sandpile''-like cellular automata such as the
forest fire model of
    \cite{BakEA90}.
The SOC paradigm has since found many applications
    \citep{BakSOC,JensenSOC,SornetteSOC},
one of which has been its use by \cite{LuHamilton91} and subsequent
authors to explain the observed power-law distributions for the
magnitudes, intensities, and durations of solar flares. SOC has since
also been applied to other natural and artificial plasma confinement
systems, notably the Earth's magnetosphere (a recent review is that of
    \cite{ChapWat01}) and tokamaks (e.g. \citep{ChapEA01}).

 In turn, recent studies in the solar flare context, of both shell
models \citep{BoffettaEA99} and simulations based more directly on the
magnetohydrodynamic (MHD) equations
\citep{GeorgoulisEA98,EinaudiVelli99} have focused attention on the
fundamental questions of the ways in which SOC and turbulence may
differ \citep{BoffettaEA99} or, conversely, the extent to which SOC
may serve as a model for turbulence \citep{EinaudiVelli99}. These more
general questions are our main focus in this paper.  Also included is some
brief
discussion of the application of these ideas to solar flares and
plasma turbulence in the solar wind, magnetosphere, and elsewhere.

Recent work
has shown
that magnetically forced 2D MHD turbulence produces power-law
PDFs in the size and duration of bursts in  spatially
averaged Ohmic energy dissipation $\langle \eta J^2 \rangle$
     \citep{GeorgoulisEA98,EinaudiVelli99} (G98).
Recall that such power-law PDFs are a necessary but not sufficient
condition for SOC
      \citep{JensenSOC}.
In addition \cite{EinaudiVelli99} showed
that a cellular automaton with
rules chosen to be consistent with the MHD model also
produces avalanches (and power-law PDFs).

In contrast, \cite{BoffettaEA99} and \cite{GiulianiEA00} (B99)
attributed the presence of a power-law PDF in the observed
waiting time between solar flares  to turbulence.
Rather than simply being due to  the scale-free turbulent cascade itself,
B99's suggested mechanism for the production
of power law waiting time PDFs was on-off intermittency.
They noted that simple prototype models
of on-off intermittency and a many-oscillator shell model of
turbulence both had waiting time PDFs which were power-law, while the
original BTW sandpile algorithm did not, having instead an exponential
PDF of waiting times.

Two important criticisms of B99's model and its interpretation have
recently been made in the context of solar flares.
    \cite{EinaudiVelli99}
pointed out that its attractors were states in which velocity and
magnetic field were aligned, whereas they asserted that force free
(aligned current and magnetic field) states were probably more
appropriate to the corona.
  \cite{Wheat00} showed how a Poisson process
varying on a long time scale (e.g. the quasi-periodic solar cycle)
could convert an exponential waiting time distribution such as that
from the original sandpile model into a power-law one of the type
observed.

These results are however  less relevant to  the more general plasma
turbulence case. In physical systems where there is no long-term
periodicity to motivate the ``periodic-Poisson" assumption of \cite{Wheat00}
there is no apparent reason to prefer such
a mechanism for observed power laws to an intrinsically scale-free
one. In addition,
the work of \cite{Wheat00} seems  to have been partly been a response
to B99's over-general assertion
that all SOC models \emph{must} have exponential waiting time PDFs.
This is not true in general (c.f. the discussions in \cite{Galtier01}
and \cite{FreemanEA00b};
and the models studied by  \cite{PaczuskiEA96}). In consequence,
several interesting questions remain open.

One is whether the burst size and duration PDFs found
in simulations of 2D MHD turbulence
     \citep{GeorgoulisEA98,EinaudiVelli99}
are also observed for either full MHD or reduced MHD
     (RMHD),
in which ``slow'' $k_z$ dependence is retained,
(see below).

The second is whether the power-law waiting time PDFs seen
 in both
the B99 one-dimensional shell model
and the \cite{Galtier01} 1D MHD simulation,
are also seen in higher dimensions.

Finally there is the question of whether a minimal set of scale-free
``burst'' PDFs and power spectra can be identified which suffices to
identify SOC (or turbulence) in a physical system. This last question
is also highly topical in magnetospheric and laboratory plasmas
\citep[cf.\ ][]{Krommes00,FreemanEA00b,KovacsEA01}.

In this paper we address these questions by examining time series of
various spatial averages of the
squared electric current density,
     $j^2(x,y,z,t)$,
drawn from a RMHD simulation.
A threshold method,
used by previous authors to construct burst-size PDFs,
is applied to the time series.
Power-laws in size (and arguably also in duration) are found, extending
the forced 2D MHD results of
     \cite{GeorgoulisEA98} and \cite{EinaudiVelli99}
to forced RMHD. The PDF of waiting times is also
not exponential, confirming that higher-dimensional RMHD is in keeping
with the predictions of B99 based on a 1D shell model. Because the
fixed threshold method employed detects fractality, which was a
predicted feature of SOC but is also generic to red noise, we then
consider to what extent the current evidence is unambiguous. We
conclude by suggesting a direction for future research.

\section{Simulation Data and Analysis}

The data analysed here is extracted from a (spectral method)
reduced MHD simulation which was used in connection with a model for
coronal heating via the coupling of low-frequency Alfv\'en waves and
quasi-2D turbulence
  (see \cite{OughtonEA01} for further details).
Using standard (nonlinear) RMHD as a base
  \citep{Mont82a,Strauss76,ZankMatt92a},  
the equations were augmented with terms representing
(i) forcing by a single large-scale Alfv\'enic mode,
(ii) reflection of all propagating modes,
and
(iii) transmission of outward propagating modes.
Physically, one may think of reduced MHD as parallel planes of
(incompressible) 2D MHD coupled together by a {\em strong\/}
mean magnetic field
  (${\bf B}_0$)
perpendicular to these planes. {\em Long\/} wavelength Alfv\'en waves
propagate along the mean field.
Thus, the fluctuating velocity and magnetic
 fields (respectively ${\bf v}$ and ${\bf b}$)
 are functions of all three spatial
coordinates, but gradients in the ${\bf B}_0$ direction are
  restricted to be weak.
Moreover, ${\bf v}$ and ${\bf b}$
are strictly perpendicular to ${\bf B}_0$.

Here we are primarily interested in various time series characterizing
such systems.
Specifically, those for the spatially averaged
 mean-square electric current density
  $ J^2 (t) = \langle j^2(x,y,z, t) \rangle / 2 $,
and the $k_z$-dependent $x$- and $y$-averaged $j^2 /2$,
denoted as
  $ J^2 (t,k_z) $,
where $k_z$ is the Fourier wavenumber in the direction parallel to the
mean field and angle brackets denote the spatial averaging.
Clearly, $ J^2(t) = \sum_{k_z} J^2(t, k_z) $.

The particular simulation employed has                   
large-scale Reynolds numbers of 800,
a resolution of $256^2 \times 4 $,
(Alfv\'enic) forcing of the ${\bf k} =  (1,1,1) $ Fourier mode,
and reflection and transmission rates of 1.0 and 0.3, respectively
    \citep{OughtonEA01}.
Note that the reflection and transmission parameters
are to be interpreted as inverse time scales and not as fractions.
The simulation was continued for 500\,$T_B$
where $T_B$ is defined as the time taken
    for a forced Alfven wave to cross the simulation box,
which is comparable to the large-scale nonlinear time.
After a few tens of $T_B$, the system settles down into a
state which is  more or less statistically steady, reminiscent of the similar
behaviour of the BTW sandpile (see figure 4.3 of \cite{JensenSOC}).
The time series are obtained by calculating the appropriate quantities
every $1/10$ of a box crossing time, after this steady state has been
reached. This sampling interval was chosen in order to have a manageable amount
of simulation data.
After removing
the initial transient, each time series $J^2(k_z, t)$ consists of
$\approx 4500$ points. The $ k_z = 1$
plane is special, since the single forced mode lies in it.

Figure 1a shows the
statistically steady portion of the time series of  $J^2(t,k_z=2)$,
for which the corresponding
PDF and cumulative distribution
function (CDF) are  shown as Figures 1b and 1c respectively.
Inspection of the  time series itself (Figure 1a) does not show extreme values
to the same extent as e.g. figure 1a of \cite{GeorgoulisEA98}, and so
in consequence the PDF we find for it (Figure 1b) is substantially more
symmetric. This is illustrated by the dashed line in Figure 1b which
shows a Gaussian with the same mean and standard deviation as those of the
time series.

To measure the distribution of bursts, a fixed threshold method was
employed, as used by \cite{FreemanEA00b} (and previous workers, see
references therein). The size $e$ of a burst was defined as the
integrated area under the curve between a given upward crossing of a
fixed threshold and the immediately subsequent downward crossing. The
duration $T$ was then the time between upward and downward crossings,
while the waiting time (or inter-burst interval, $\tau$) was that
between a given downward crossing and the next upward crossing.  The
resulting PDF for burst size is plotted in Figure~2a, where the solid
line indicates the curve corresponding to use of the median value of
the time series as the threshold, while the eight dotted lines
correspond to those resulting from the 10th, 20th, \ldots, 40th, 60th
through 90th percentiles. PDFs for burst duration and waiting time
constructed by the same method are shown in Figures~2b and 2c.
The number of bursts
bursts thus defined will vary weakly with the threshold chosen, in the case
of the median threshold the size distribution $D(e)$ was formed from
198 events.
Despite the symmetrical PDF of $J^2$ shown in Figure 1b,
a power-law PDF is obtained for burst sizes in the range $10^{-3}$
to 2 units (Figure~2a),
which remains stable even as the thresholds are varied.
Outside this region the points deviate from a power law but their statistical
weight is low. The dashed line shown is a power law fit to those points
where the number of samples per bin is greater than 4.
 The PDFs of burst durations and waiting times also
resemble power-laws
in the range 0.2 to 2, beyond which the number of samples per bin again
falls below 5.
Similar plots were obtained for all four $k_z$ planes in the simulation,
and on averaging over $k_z$.
Figure 2d shows the PDF for the waiting times
plotted on semilog axes, on which an exponential distribution would
appear as a straight line, confirming that the waiting times (as
defined herein) for this simulation
are {\it not\/} exponentially distributed.

\section{Discussion and Conclusions}

We find that the PDF of burst size
measured in our simulation is  power-law in form,
independent of the choice of threshold.
The PDFs of duration and waiting time appear to have the same
basic form (although the evidence is
much less clear cut). Intriguingly, we have observed this
property  in a time series (Figure 1a) whose PDF
(Figure 1b)  is not long-tailed but rather symmetric.
Power-law burst PDFs are often seen coupled with
long-tailed underlying PDFs, and so it is sometimes
thought that the scaling range of a signal will be controlled by the
mean to peak ratio (governed by $\sigma$ and $\mu$ for Gaussian data,
$\mu$ for Poisson data, etc.)  In fact, however, burst size is also
governed by the degree of persistence $\beta$ in the signal, because
it is not just affected by the probability of N points being above a
line but is controlled by the probability of N non-independent,
successive points all being above the line.
  \cite{MalTurc99}
have noted that $\beta$ can be varied
independently of the PDF of the members of a time series, so that even
a time series with a Gaussian distribution of amplitudes but a
non-zero $\beta$ would have a non-negligible probability of several
successive values exceeding a threshold. In addition, for such a
series, the scaling properties of the distributions of waiting time
and inter-burst interval are set entirely by $\beta$ and would be
power laws.

This raises an interesting question, however  (see also \cite{FreemanEA00b}).
Red noise (``$1/f$'') time series of the type which
SOC systems were originally expected to produce, and whose appearance
SOC was proposed to account for, are fractal. The distribution of
isosets\footnote{Defined
     as the set of times at which the time series
     crosses a fixed level.}
of such a time series is a power-law \citep{AddisonFractals}. Hence
the burst duration and waiting time PDFs of red noise when
found by a threshold method should be power-laws, regardless of
whether the noise is produced by turbulence or one of the class of
SOC-type models which do produce red noise.
The original BTW model can be
eliminated as its waiting time series was later shown to be
uncorrelated and thus not red noise
  \citep{JensenSOC,BoffettaEA99,FreemanEA00b}.
To determine if a process is SOC in the sense of BTW's original
proposal, one needs information about spatial correlations as well as
temporal correlations.  This is because SOC was proposed as a
mechanism linking spatial fractality and temporal persistence (``1/f"
noise).  This reinforces the point that, rather than simply temporal
information (burst durations and waiting times) or avalanche
distributions (``burst sizes"), to differentiate between turbulence
and SOC it will be necessary to make unambiguous predictions about
spatial structure for each phenomenon, requiring at minimum then
availability of spatial correlation functions. We note that
correlation between bursts e.g. the debated ``sympathetic" quality of
solar flares \citep{WheatEA98} could in principle occur through time
or space correlation.  This line of investigation will be pursued in
future papers.

\section{Acknowledgements}
Support for SO was provided by grants from
NASA (SECTP NAG5-8134)
and PPARC (PPA/G/S/1999/00059). The RMHD simulations were performed on
VAX Alphas at the Bartol Research Institute.
NWW acknowledges valuable discussions with
S\'ebastien Galtier, Bruce Malamud, Richard Dendy, Sandra Chapman, Per Bak
and Maya Paczuski.

{\scriptsize
  \newcommand{\SortNoop}[1] {}
  \newcommand{\MHD}  {{M}{H}{D}\ }
  \newcommand{\mhd}  {{M}{H}{D}\ }
  \newcommand{\wkb}  {{W}{K}{B}\ }
  \newcommand{\alfven} {{A}lfv\'en\ }
  \newcommand{\Alfven} {{A}lfv\'en\ }
  \newcommand{\alfvenic} {{A}lfv\'enic\ }
  \newcommand{\Alfvenic} {{A}lfv\'enic\ }

}

\newpage
\begin{figure}
\centering
\psfig{figure=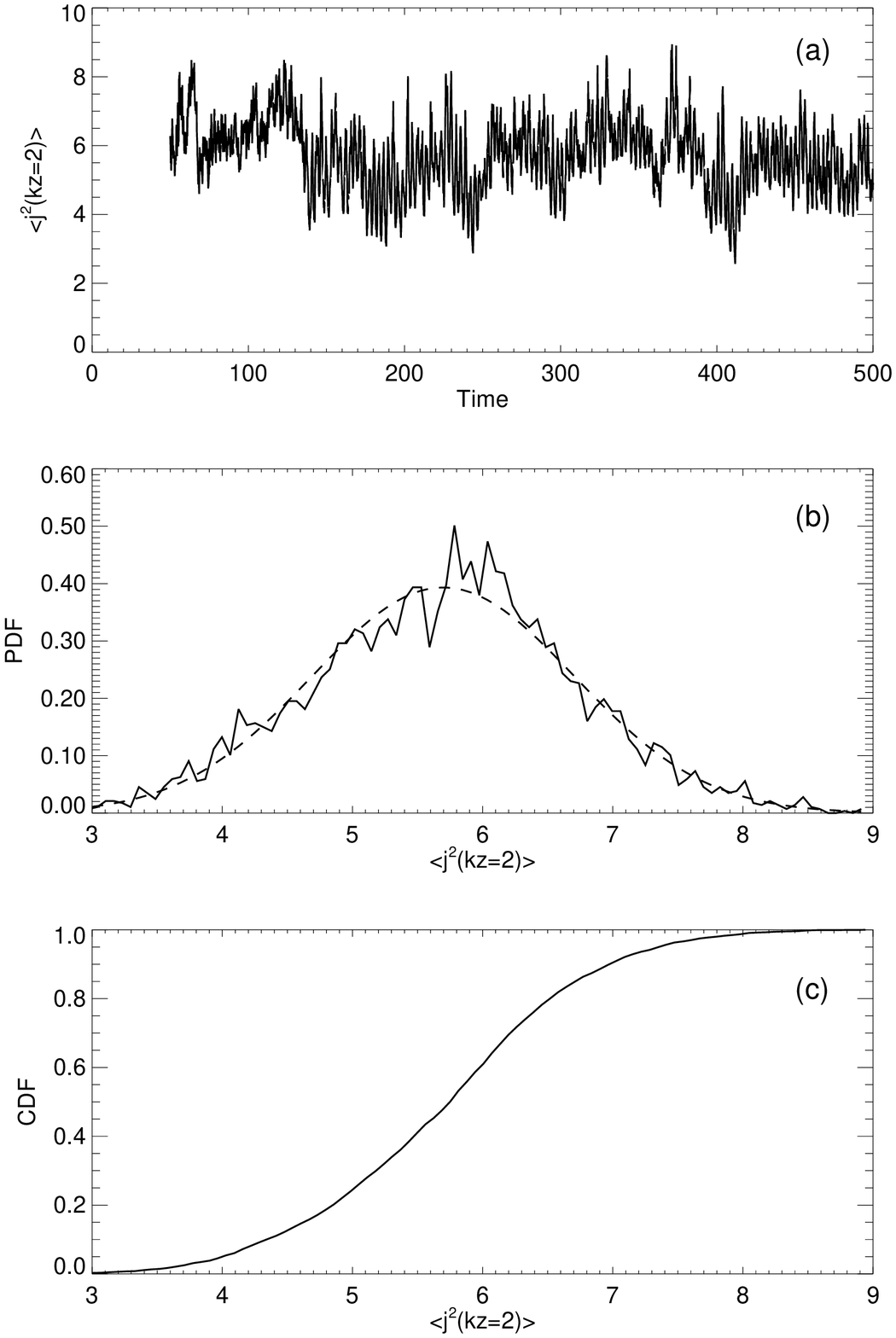,width=.9\textwidth}
\caption{ a) Time series of current density for the reduced MHD simulation.
b) PDF of current density for the time series in a). Overplotted is a Gaussian
distribution with the mean and standard deviation of the time series in Figure 1a. c) CDF of current density
for the time series in a).
} \end{figure}

\begin{figure}
\centering
\psfig{figure=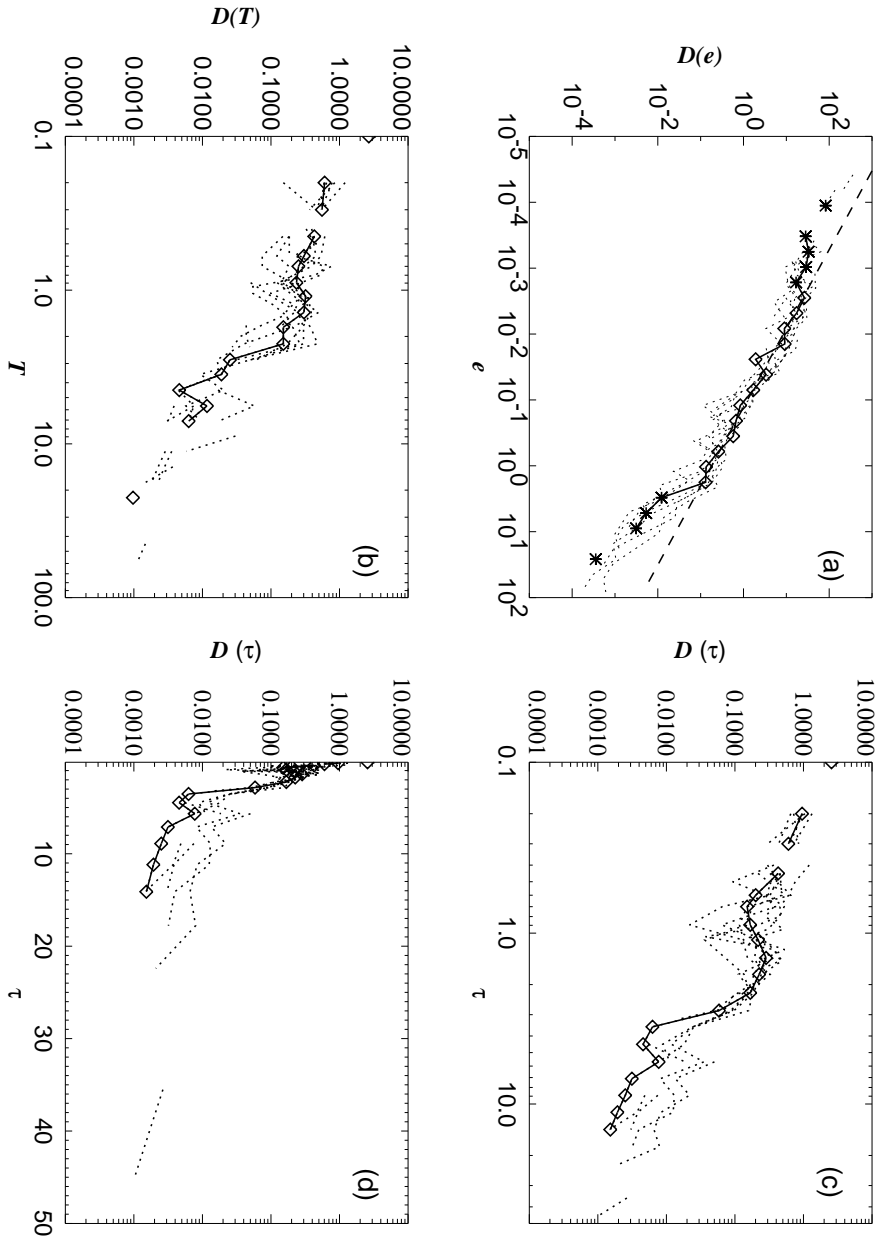,width=.9\textwidth}
\caption{ a) Burst size PDF obtained by the threshold method
for the time series of Figure 1a). The solid line corresponds to the use of the median of
the time series as the threshold while the dotted lines show the 10th, 20th, $\dots$ 90th
percentiles. The dashed line passes through those points for which a statistically
significant number of points is available. b) Burst duration PDF obtained by the method of Figure 2a). c) PDF of
waiting times between bursts obtained by the method of Figure 2a). d) The PDF
of figure 2c) plotted on a log-linear scale, illustrating that the PDF cannot be
fitted well by an exponential.
} \end{figure}

\end{document}